\documentclass[12pt,a4paper]{cibb}

\makeatletter
\providecommand{\@ordinalM}[2]{#1}
\makeatother

\usepackage{subfigure,graphicx}
\usepackage{amsmath,amsfonts,latexsym,amssymb,euscript,xr}
\usepackage{booktabs}
\usepackage[nodayofweek]{datetime}
\usepackage{hyperref}
\usepackage{fmtcount}
\usepackage[english]{datenumber}
\usepackage[absolute]{textpos}
\usepackage{enumitem}
\usepackage{tabularx}
\usepackage{array}
\usepackage{lscape}

\usepackage[table]{xcolor}
\usepackage{color,colortbl,tabularx}

\usepackage[english]{babel}
\usepackage[protrusion=true,expansion=true]{microtype}
\usepackage{amsmath,amsfonts,amsthm}
\usepackage{pifont}

\def\black{\color{black}}
\def\blue{\color{blue}}

\definecolor{LightBlue}{rgb}{0.88,0.9,0.9}

\title{\Large $\ $\\ \bf A Computational Ethical Framework for Financial Digital Phenotyping for Mental Health}

\author{\blue \large Oluwadara Adedeji$^{*1}$, Michael Mayowa Farayola$^{2}$, Jeff Brozena$^{3}$, \\
Irina Tal$^{2}$, Regina Connolly$^{4}$, and Mark Matthews$^{1}$ }
\address{\blue \footnotesize $\ $\\$^1$ School of Computer Science, University College Dublin, Dublin, Ireland. \\
$^2$ School of Computing, Dublin City University, Dublin, Ireland. \\
$^3$ College of Information Sciences and Technology, Pennsylvania State University,University Park, PA, USA.  \\
$^4$ Business School, Dublin City University, Dublin, Ireland. \\
\bigskip
ORCID codes: OA 0000-0002-3430-6011; MMF 0009-0002-3495-4155; JB 0000-0001-7621-1566; \\ IT 0000-0001-9656-668X; RC 0000-0003-3196-2889; MM 0000-0001-8862-5141.
\bigskip
\newline
$^*$corresponding author: oluwadara.adedeji@ucdconnect.ie
}

\abstract{\small Computational ethics, digital phenotyping, financial data, financial technologies. \normalsize
\\[17pt]
{\bf Abstract }
Ethical governance of AI-driven systems is often expressed through high-level principles and static documentation, creating a gap between regulatory requirements and system-level verification. This challenge is particularly acute in digital phenotyping, where continuous behavioural data raises concerns around consent, privacy, and fairness. In this paper, we propose a computational ethical framework for AI-driven digital phenotyping system in which ethical requirements are formalised as deontic temporal logic constraints, alongside a conceptual ethical agent that oversees the system and ensures that any supervised system satisfies the specified constraints. Using a case study involving financial data and mental health, we model key ethical properties and verify them using the Z3 Satisfiability Modulo Theories (SMT) solver. Our evaluation shows that the framework is logically consistent and that violations of the specified ethical properties are ruled out within the formal model through counterexample-based verification. This presents early research enabling continuous, machine-verifiable ethical checking, moving beyond retrospective compliance based on static documentation. We discuss limitations, including the need for real-world verification with data, the challenge with subjectivity and contextual sensitivity, the need for human oversight, and outline how such approaches can support the development of digital phenotyping and AI systems with continuous and auditable ethical guarantees.

}

\begin{document}

\renewcommand{\thefootnote}{}
\footnotetext{\small{Article version: \datedate $\;$ h\currenttime  $\;$ CET}}

\thispagestyle{myheadings}
\pagestyle{myheadings}
\markright{\tt Proceedings of CIBB 2026}

\section{Introduction}
\label{sec:SCIENTIFIC-BACKGROUND}

Digital phenotyping, the moment-by-moment quantification of individual behaviour using data from personal devices such as smartphones and wearables as digital biomarkers~\cite{onnelaHarnessingSmartphoneBasedDigital2016}, has emerged as a promising approach for passive monitoring and early detection of mental illness. These systems increasingly rely on artificial intelligence (AI) to infer behavioural and clinical states from continuous data streams, introducing new challenges related to privacy, consent, fairness, and data governance that limit their clinical adoption.

While these ethical challenges are well studied, they are typically expressed as high-level principles rather than formal specifications that can be directly implemented or verified within AI systems. Prior work identifies a range of ethical challenges in digital phenotyping, including issues related to data provenance, algorithmic bias, transparency, and accountability~\cite{mulvennaEthicalIssuesDemocratizing2021,martinez-martinEthicalDevelopmentDigital2021}, as well as difficulties in obtaining meaningful consent in vulnerable populations~\cite{schweigerEthicsAIassistedDigital2025}. However, these approaches do not provide computational mechanisms to oversee and enforce such requirements in AI systems~\cite{gorelikEthicsAIHealthcare2025}. As a result, there remains a gap between ethical requirements and actual system behaviour, particularly in settings where decisions are automated and continuously evolving.

Addressing this gap requires moving ethics from high-level principles and static requirements to computationally verifiable mechanisms. In this work, we adopt a computational ethics approach in which ethical requirements are formalised as machine-interpretable constraints. We argue that AI agents, software components that can monitor and act on system behaviour, provide a natural mechanism for operationalising such constraints. Specifically, an ethical oversight agent can be integrated into a digital phenotyping system to continuously monitor system behaviour and verify compliance with specified ethical requirements, with the potential to support enforcement. This enables a shift from static documentation to machine-verifiable checking and oversight. Formal methods, including deontic and temporal logic, support the specification of obligations and prohibitions over time, while constraint-solving techniques enable verification and detection of violations~\cite{awadComputationalEthics2022}. Despite this potential, agent-based approaches for overseeing and enforcing formalised ethical constraints remain largely unexplored in digital phenotyping.

To instantiate this framework, we focus on financial data as an emerging modality for digital phenotyping~\cite{blairFinancialTechnologiesFinTech2022}. Financial behaviour has been shown to reflect mental health states, for example through increased spending during manic or hypomanic episodes in bipolar disorder~\cite{richardsonRelationshipBipolarDisorder2017}. This modality introduces distinct characteristics and challenges: financial data is highly sensitive, and its misuse carries direct legal and economic consequences, raising ethical concerns beyond those typically associated with smartphone data; it enables passive, scalable collection of behavioural signals with minimal user burden, potentially reducing missing data caused by dropout in smartphone- and wearable-based studies; and ethical considerations and formal frameworks for this data modality are largely absent from the literature, highlighting a critical gap in digital phenotyping ethics. 

From a regulatory perspective, such systems are classified as high-risk under the EU Artificial Intelligence Act~\cite{eu_ai_act}, which imposes requirements including data governance, fairness and non-discrimination and human oversight. However, these requirements are specified at a high level, creating a gap between regulatory obligations and technical verification.

In this paper, we propose a computational ethical framework for financial digital phenotyping in which ethical requirements are encoded as deontic temporal logic constraints and evaluated by a conceptual ethical oversight agent. We use the Z3 Satisfiability Modulo Theories (SMT) solver to verify consistency and violations.
The framework supports hybrid human–AI oversight through continuous ethical evaluation, analogous to conformance testing in high-stakes domains.

\section{Methods}
\label{sec:METHODS}

Deontic logic is a formal framework for reasoning about normative concepts such as obligation $(O\varphi)$, permission $(P\varphi)$, and prohibition $(F\varphi)$. Deontic temporal logic extends this framework with temporal operators (e.g., $G$ for \emph{always}) to model these norms over time. Giordano et al. \cite{giordanoTemporalDeonticAction2013} contain expanded lessons on mathematical deontic logic formulations. Note that prohibition is logically equivalent to the obligation of the 
negation: $F(\varphi) \equiv O(\neg\varphi)$.

We identified ethical issues through our institutional ethics review and Data Protection Impact Assessment (DPIA), a structured process for assessing and mitigating risks associated with data processing in our ongoing financial digital phenotyping study. This was complemented by a review of challenges in the literature and the EU AI Act, particularly in data collection and modelling. While a typical AI pipeline includes stages such as deployment and monitoring, we focus on data collection and modelling, as these remain the primary stages in current digital phenotyping research, especially for financial data.


We model ethical requirements using deontic temporal logic to represent obligations and prohibitions over time-dependent system behaviour. These deontic constructs are compiled into logical constraints for verification. We define a conceptual ethical agent that supervises the financial digital phenotyping system and ensures that any supervised system satisfies the formalised ethical constraints. Predicates are defined to represent system states, and used within deontic expressions to encode ethical rules. We then assess satisfiability and detect potential conflicts using the Z3 solver.


We define the domains, variables, and predicate symbols used to formalise the ethical constraints of the agent in this study as follows:

\vspace{-8pt}

\begin{align*}
&\textbf{Domain (sort) Variables}\\[2pt]
p &\in \mathcal{P} :\ \text{set of participants} \\
t &\in \mathcal{T} :\ \text{time domain} \\
f &\in \mathcal{F} :\ \text{set of financial data fields} \\
\rho &\in \mathcal{R} :\ \text{set of specified research purposes} \\
A &\in \mathcal{A} :\ \text{set of agents} \\
S &\in \mathcal{S} :\ \text{set of digital phenotyping systems} \\[4pt]
&\textbf{Predicates}\\[2pt]
\text{Cap}(p) &:\ \text{Participant } p \text{ has decision-making capacity} \\
\text{Info}(p) &:\ \text{Participant } p \text{ is adequately informed} \\
\text{Vol}(p) &:\ \text{Consent from participant } p \text{ is voluntary} \\
\text{Withdraw}(S,p,t) &:\ \text{System } S \text{ records withdrawal of participant } p \text{ at time } t \\
\text{ConsentValid}(p,t) &:\ \text{Consent of participant } p \text{ is valid at time } t \\
\text{UseData}(S,p,t) &:\ \text{System } S \text{ uses data from participant } p \text{ at time } t \\
\text{Collect}(S,f,p,t) &:\ \text{System } S \text{ collects field } f \text{ from participant } p \text{ at time } t \\
\text{IdentifyingField}(f) &:\ \text{Field } f \text{ is identifying or linkable} \\
\text{Necessary}(f,\rho) &:\ \text{Field } f \text{ is necessary for research purpose } \rho \\[4pt]
\text{Fair}(S) &:\ \text{System } S \text{ satisfies a defined fairness criterion} \\
\text{BiasMitigated}(S) &:\ \text{Bias in system } S \text{ is mitigated} \\
\text{Discriminate}(S,p,t) &:\ \text{System } S \text{ produces a discriminatory outcome for participant } p \text{ at time } t \\
\text{Responsible}(S) &:\ \text{System } S \text{ satisfies responsibility criteria} \\
\text{Supervises}(A,S) &:\ \text{Agent } A \text{ supervises system } S \\
\text{Ethical}(A) &:\ \text{Agent } A \text{ satisfies all ethical constraints}
\end{align*}


\section{Results}

\label{sec:RESULTS}
\subsection{\bf \it Deontic Temporal Logic for Financial Digital Phenotyping}
\label{sec:deontic-logic}

To formulate and verify the ethics of this system logically, 
we represent the properties as follows, where $S$ denotes 
the financial digital phenotyping system and $A$ denotes 
the conceptual ethical oversight agent:

\subsubsection{\bf \it Proposition for Financial Digital Phenotyping}
\begin{enumerate}[label=\textbf{\alph*)}, topsep=2pt, itemsep=2pt, parsep=0pt]

\item Informed consent is valid only if a participant has the capacity to make the decision, is provided with clear and sufficient information about the study, gives consent voluntarily, and retains the right to withdraw that consent at any time without penalty.
\vspace{-10pt}

\begin{align}
O\big(\text{ConsentValid}(p,t) &\rightarrow (\text{Cap}(p) \land \text{Info}(p) \land \text{Vol}(p))\big) \\
O\big(\text{Withdraw}(S,p,t) &\rightarrow G_{t'>t}\ \neg \text{ConsentValid}(p,t')\big)
\end{align}
\vspace{-10pt}

These constraints specify obligations on the system. First, informed consent is considered valid only if, for a participant at time $t$, the participant must have the capacity to consent, be adequately informed, and act voluntarily. These conditions may be operationalised through AI or human assessments of capacity, clarity of information provided, and measures of voluntariness (e.g., questionnaires). Second, if the system records a withdrawal initiated by a participant at time $t$, then consent must be invalidated for all future times $t' > t$. In (\textbf{b}), this implies that the system must cease any further data collection or use once consent is withdrawn, for example by deleting or disabling access to the participant’s data. Note that “only if” denotes a necessary condition, not a sufficient one.

\item Data must not be collected or used if informed consent is not valid.
\vspace{-15pt}

\begin{align}
F\big((\text{Collect}(S,f,p,t)\lor \text{UseData}(S,p,t)) \land \neg \text{ConsentValid}(p,t)\big)
\end{align}

\vspace{-3pt}

Data collection and data use are modeled as separate predicates, allowing for cases where data is used without being collected by the system. The disjunctive condition ensures that consent is required whenever either operation occurs.

\item It is forbidden to collect identifying financial data fields unless they are necessary for the specified research purpose.
\vspace{-6pt}
\begin{align}
F\big(\text{Collect}(S,f,p,t)\land \text{IdentifyingField}(f)\land \neg \text{Necessary}(f,\rho)\big)
\end{align}
\vspace{-15pt}

This prevents the collection of fields, such as merchant names, that contain identifiable information without a clear research need.

\item If the system is responsible, then it must satisfy fairness constraints, mitigate measurable biases, and not produce discriminatory outcomes. 
\vspace{-6pt}
\begin{align}
O\Big(
\text{Responsible}(S) \rightarrow \text{Fair}(S) \land \text{BiasMitigated}(S) \land \forall p,t\ \neg\text{Discriminate}(S, p, t)
\Big)
\end{align}
Fairness ensures equitable outcomes across individuals or groups, bias mitigation addresses systematic distortions in data or models, and discrimination refers to unjust or unequal treatment resulting from such biases. These conditions are necessary for responsibility, and a system cannot be considered responsible without satisfying them. Additional criteria, such as transparency and accountability, may further 
constrain responsibility in practice but are not formalised here.

\end{enumerate}

\noindent
\textbf{Explicit Ethical Agent}: An agent $A$ is considered ethical only if every system it supervises satisfies all specified ethical constraints ((a)–(d))
\vspace{-5pt}
\begin{align}
\text{Ethical}(A) \rightarrow
\forall S \in \mathcal{S}\,
\Big(
\text{Supervises}(A,S) \rightarrow
\text{Constraints}(S)
\Big)
\end{align}
\vspace{-25pt}
\begin{align}
\text{Constraints}(S) \;\triangleq\;
\text{Eq.}(1) \land \text{Eq.}(2) \land \text{Eq.}(3) \land \text{Eq.}(4) \land \text{Eq.}(5)
\end{align}
\vspace{-10pt}

The consent constraints (Eqs.~(1)--(3)) correspond to the data governance obligations of Article 10 of the EU AI Act, while the responsibility constraint (Eq.~(5)), which subsumes fairness and non-discrimination, operationalises the bias mitigation requirements of Article 10. This demonstrates that the high-level obligations of the EU AI Act can be operationalised as machine-verifiable constraints, providing a pathway toward continuous, auditable compliance rather than reliance on static documentation alone.

\subsection{\bf \it Verification}
\label{sec:verification}

We evaluated the formal model using the Z3 SMT solver for both global consistency and scenario-based verification. Eqs. (1)–(5) are asserted as global constraints, while the ethical-agent condition (Eq. 6) defines an agent as ethical only if every system it supervises satisfies all specified ethical constraints.


The global satisfiability check returned \textit{sat}, indicating that 
Eqs.~(1)--(6) are logically consistent and that at least one assignment 
of variables and predicates satisfies all constraints. However, the 
generated satisfying model is vacuous, as several key system behaviours, 
including data collection, data use, withdrawal, and supervision, are 
assigned \textit{False} by the solver. This is expected because the global 
satisfiability check verifies logical consistency rather than realistic 
system behaviour. Therefore, meaningful behaviours are evaluated separately 
through counterexample-based and positive verification tests.


We evaluated the framework using counterexample-based verification by formulating a violating scenario for each ethical property and submitting it to the solver. These scenarios included invalid informed consent, continued consent following withdrawal, data collection or use without valid consent, unnecessary collection of identifying financial data, violations of responsible AI requirements, and an ethical oversight agent supervising a non-compliant system. In every case, the solver returned \textit{unsatisfiable}, confirming that the proposed constraints prevent the corresponding ethical violation. Furthermore, a positive test confirmed that the framework permits data collection when all ethical requirements are satisfied.



\section{Discussion}
\label{sec:discussion}
The proposed framework represents an initial step toward formalising 
computational ethics in digital phenotyping. By translating ethical 
requirements into logical constraints and verifying them using the Z3 
solver, the framework moves ethical governance from abstract principles 
toward an operational and testable approach. Rather than replacing human 
oversight, it is designed to complement existing ethical review processes 
through continuous and systematic verification. This human-in-the-loop 
approach, aligned with Article~14 of the EU AI Act, preserves human 
judgement while supporting more robust ethical governance. It may also 
enable controlled ``sandbox'' environments where systems can be evaluated 
for ethical compliance before and after deployment. The conceptual ethical 
oversight agent could be implemented as an AI agent powered by existing 
large language models (LLMs), with the proposed ethical rules and constraints 
integrated into its reasoning and decision-making processes.

While the constraint system is logically consistent, its ethical guarantees are better validated through targeted scenario-based checks than through global satisfiability alone. While we tested counterexamples within the formal model, further evaluation with real-world data and systems may reveal additional scenarios and strengthen the assessment of the framework’s robustness.

However, several limitations remain. First, the model assumes that concepts 
such as capacity, informed consent, and voluntariness can be formally defined 
and verified. In practice, these concepts are context-dependent and often 
require human interpretation, particularly in mental health settings. For 
example, assessing a participant's capacity to provide informed consent may 
require clinical judgement, while evaluating voluntariness may depend on 
individual and social contexts. Although hybrid human--AI approaches could 
support these assessments by escalating uncertain cases for human review, 
further research is needed to formally represent these complex concepts. 
Second, the framework focuses on consent-based data processing and does not 
currently capture alternative legal or ethical bases, such as proxy 
decision-making, public interest, or clinical necessity.

Third, the use of formal logic introduces limitations in handling ambiguity, uncertainty, and context-dependent scenarios, indicating the need for hybrid approaches such as neurosymbolic methods that combine rule-based reasoning with data-driven models. While the framework is aligned with the EU AI Act, it does not fully account for variations in ethical norms across different regional, cultural, or institutional contexts.  Its scope is also limited to data collection and modelling, excluding important considerations related to deployment, monitoring, data retention, and lifecycle management.

In addition, the framework does not evaluate whether the underlying research purpose (i.e., $\rho$) is itself ethically justified. Adversarial or unethical uses, such as attempts to manipulate or bypass constraints, may undermine automated ethical verification. More broadly, the model does not capture socio-technical system (STS) dynamics, where ethical outcomes depend not only on system design but also on real-world use. For example, financial-data–based mental health assessments could be misused for discriminatory decisions in insurance or credit contexts, even if the model itself satisfies formal fairness constraints. 

Despite these limitations, the proposed framework demonstrates that 
high-level obligations of the EU AI Act can be operationalised as 
machine-verifiable deontic constraints, providing a computational foundation 
for formalising ethical requirements in digital phenotyping. This supports 
a shift from reliance on static documentation toward more transparent, 
auditable, and verifiable approaches to ethical compliance. As the need for 
practical compliance mechanisms in digital health continues to grow, this 
framework represents a step toward regulatory-ready systems in domains where 
privacy, trust, and fundamental rights are critical considerations.

\section{Conclusion}
\label{sec:CONCLUSIONS}

This study proposes a computational ethical framework for financial digital phenotyping for mental health, in which ethical requirements are formalised as deontic temporal logic constraints and verified using the Z3 SMT solver. We introduce a conceptual ethical oversight AI agent that supervises financial digital phenotyping systems to ensure compliance with key ethical requirements, including informed consent, data minimisation, fairness, and non-discrimination. Counterexample-based verification demonstrates that these ethical properties can be systematically verified, providing a foundation for continuous, machine-verifiable governance of AI systems that use objective financial data to support mental health assessment. To our knowledge, this is the first work to apply computational ethics methods to digital phenotyping, offering a pathway towards trustworthy, privacy-preserving, and regulatory-aligned AI for mental health applications.

Future research should evaluate the framework in real-world mental health digital phenotyping systems, particularly those using objective behavioural data such as financial transactions to support early detection and monitoring of conditions such as bipolar disorder. Extending the framework to incorporate additional legal and ethical bases beyond consent, model the full AI lifecycle, address ambiguity through neurosymbolic approaches, and account for socio-technical and cultural factors will further strengthen its applicability. These advances will support the development of safe, trustworthy, and ethically governed AI technologies for mental healthcare.

\section*{Conflict of interests}
\label{sec:CONFLICT-OF-INTERESTS}
The authors declare no conflict of interest.


\section*{Funding}
\label{sec:FUNDING}

\black
This publication is the result of research conducted with the financial support of Health Rhythms Inc.\ and Taighde Éireann -- Research Ireland under Grant Nos.\ 18/CRT/6183 (ML-Labs), 13/RC/2094\_P2 (Lero), and 13/RC/2106\_P2 (ADAPT), and is co-funded by the European Regional Development Fund (ERDF). This work is also based upon work supported by the National Science Foundation Graduate Research Fellowship Program under Grant No. DGE1255832. Any opinions, findings, and conclusions or recommendations expressed in this material are those of the author(s) and do not necessarily reflect the views of the National Science Foundation.


\section*{Availability of data and software code}
\label{sec:AVAILABILITY}
The software implementation supporting this work is publicly available at:
\url{https://github.com/darasiemi/ethics-cibb-2026}

\footnotesize
\bibliographystyle{unsrt}
\bibliography{references.bib} 
\normalsize

\end{document}